\documentclass{ws-procs975x65}

\begin{document}

\title{Ultra High Energy Cosmic Rays}

\author{Pasquale Blasi}

\address{INAF/Osservatorio Astrofisico di Arcetri, Largo E. Fermi, 
5 Firenze (Italy)\\E-mail: blasi@arcetri.astro.it}

\maketitle

\abstracts{ 
The detection of cosmic rays with energy around and in excess of 
$10^{20}$ eV raises many questions that future experiments will 
help answering to. I address here my view of some of these open 
issues, as they are now and as they might be affected by future 
observations.
}

\section{Introduction}

In the eighty years of history of cosmic rays, there has been a constant
search for the end of the cosmic ray spectrum. It has long been thought 
that this end of the spectrum would be determined by the highest energy 
that cosmic accelerators might be able to achieve. Despite this
continuous search, no end of the spectrum was found. In 1966, right after 
the discovery of the cosmic microwave background (CMB), it was understood 
\cite{GZK66} that high energy protons would inelastically 
scatter the photons of the CMB and produce pions. For homogeneously distributed
sources this would cause a flux suppression, called the {\it GZK cutoff}: for 
the first time the end of the cosmic ray spectrum was related to a physical 
process rather than to speculations on the nature of the accelerators. 
Moreover, for the first time, the end of the cosmic ray spectrum was predicted 
to be at a rather well defined energy, around $10^{20}$ eV, where the so-called
photopion production starts to be kinematically allowed. Forty years later, 
we are still seeking a confirmation that the cosmic ray spectrum has in fact 
such a flux suppression.
The two largest experiments currently operating in the energy range of 
interest, namely AGASA \cite{AGASA} and HiRes \cite{HIRES}, 
appear to have discrepant results in the 
highest energy end of the spectrum. While the data collected by the former 
appear to be consistent with the extension of the lower energy spectrum, the
latter experiment suggests that the GZK feature may be present in the data.
The small statistics of events however does not allow to draw a definitive
conclusion about the detection of the GZK feature \cite{daniel1}. A positive
detection would be a proof of the extragalactic origin of UHECRs.

The small statistics of events also affects our ability to assess the
significance of another piece of information, that comes from the AGASA
experiment: some of the events are clustered on angular scales comparable
with the angular resolution of AGASA \cite{agasaclusters}. If confirmed, 
this would represent the first true indication that UHECRs are in fact 
accelerated in some still unknown powerful astrophysical objects. 

Much progress has been made in the understanding of the physics behind
the acceleration processes that may be able to energize cosmic rays to
the highest observed energies. This is particularly true for shock 
acceleration, that still remains the most promising candidate acceleration
process, probably because it is the one that has been studied the most.

In this short review, after summarizing the main observational facts 
(Section \ref{sec:obs}), I will discuss the main open questions that
seek an answer (Section \ref{sec:open}). 

\section{A short summary of the observations}
\label{sec:obs}

In this section I briefly summarize the main pieces of the puzzle of UHECRs:
\par\noindent
{\it Isotropy}: The directions of arrival of the events at energies above 
$\sim 4\times 10^{19}$ eV appear isotropically distributed in the sky.
No immediate association with local structures (galactic disc, supergalactic
plane) arises from the data.\par
\noindent
{\it Lack of source identification}: No association of the observed events 
with known powerful nearby sources has been found. This may represent a serious
problem for the highest energy events, with energy higher than $10^{20}$ eV, 
for which the loss length is small and the sources are forced to be closeby. 
However, even at $4\times 10^{19}$ eV the loss length becomes of the same order
of magnitude of the size of the universe and it is therefore difficult to find 
a counterpart, in particular because of the poor angular resolution of current 
experiments.\par
\noindent
{\it Small Scale Anisotropies (SSA)}: The AGASA data show several doublets and
triplets of events on angular scales comparable with the resolution of
the instrument \cite{agasaclusters}. 
The statistical significance of these multiplets is still 
the subject of some debate [see for instance \cite{finley2003}], but if 
confirmed as not just the result of statistical fluctuations they 
could in fact represent the first evidence that UHECRs are accelerated 
in astrophysical point sources. 
This evidence would point against most so-called top-down models, in which 
the emission is truly diffuse.\par
\noindent
{\it The composition}: At the highest energies the information about the
chemical composition is so far very poor. A reanalysis of the Haverah 
Park inclined showers allowed to constrain the fraction of gamma rays
at energy larger than $4\times 10^{19}$ eV to about $50\%$ \cite{Ave2002}
(see the contribution of Alan Watson in these proceedings \cite{watson}).

\section{Open questions}
\label{sec:open}

In this section I address some questions that seem to me as particularly 
important for the understanding of the origin of UHECRs. The list is not 
supposed to be complete and should only be considered as a possible starting 
point for further investigations.

\subsection {Should we expect a sharp GZK cutoff?}

For a long time, the issue of detecting the GZK flux suppression has been 
interpreted as the search for cosmic ray events with energy above $10^{20}$
eV and of nearby sources of these particles. It is often forgotten that the
GZK feature is not avoided at all if the source density is constant 
(homogeneous distribution of the sources) all over the universe. As already 
mentioned above, a flux suppression is expected due to the fact that 
above the threshold for photopion production protons lose energy rapidly.
\begin{figure}[ht]
\epsfxsize=11cm   %width of figure - will enlarge/reduce the figures
\epsfbox{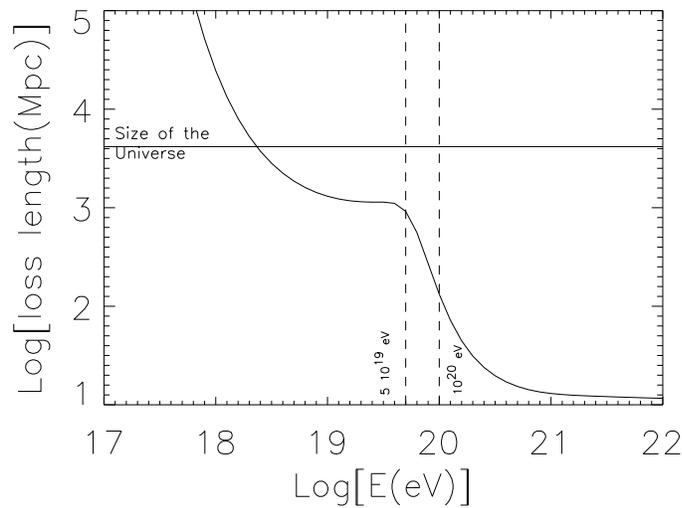}
%\figurebox{2cm}{3cm}{} %to have a box alone 
%\centerline{\epsfxsize=3.9in\epsfbox{procs-fig1.eps}}   
\caption{Loss length of UHECRs for proton pair production and photopion
production.}
\label{fig:loss}
\end{figure}
It is worth stressing however that what has been named the GZK cutoff is 
in fact a {\it feature} that can be more or less pronounced depending upon 
details such as the injection spectrum of cosmic rays, the luminosity evolution
of the sources, the local overdensity of sources and the magnetic field 
strength in the intergalactic medium. In general terms, the GZK feature is 
not determined by the number of events above $10^{20}$ eV but rather by the 
relative fraction of events above $10^{20}$ eV compared with that at 
lower energies. This is easily explained through Fig. \ref{fig:loss}, where 
I plot the loss length for proton pair production and photopion production 
(solid curve) as a function of energy\cite{bere}. The straight horizontal line 
represents the size of the universe today, while the dashed lines are just 
there to drive the eye to identify the energies of $10^{20}$ eV and 
$5\times 10^{19}$ eV. This change by a factor of 2 in the energy changes the
loss length by almost one order of magnitude, which translates approximately
into the same ratio between the flux below and above 
$5\times 10^{19}$ eV if the sources have no luminosity evolution and no
local overdensity and there is no magnetic field. A luminosity evolution 
that makes sources at high redshift brighter than nearby sources enhances
this jump, making the feature more pronounced. A local overdensity of 
sources on the other hand, has the opposite effect. Moreover, hard spectra
contain more particles in the high energy part, therefore they correspond
to a GZK feature which is less pronounced than that due to a soft injection
spectrum. All these effects were discussed at length for the first time in 
\cite{beregrigo}.

In addition to all these effects, there are some observational issues to
take into account: all experiments currently operating have a $\sim 30\%$
statistical error in the energy determination and future experiments 
will probably not make a big improvement in this respect. Since the 
spectrum of cosmic rays is steep, there is a larger number of events 
that are given a higher energy than viceversa, which makes the observed GZK 
feature look smoother. 
The effect of a systematic error in the energy determination 
clearly depends on which way it goes: as shown in \cite{daniel1}, such an 
error can have important implications in the comparison between different 
experiments. 

\begin{figure}[ht]
\epsfxsize=6cm   %width of figure - will enlarge/reduce the figures
\epsfbox{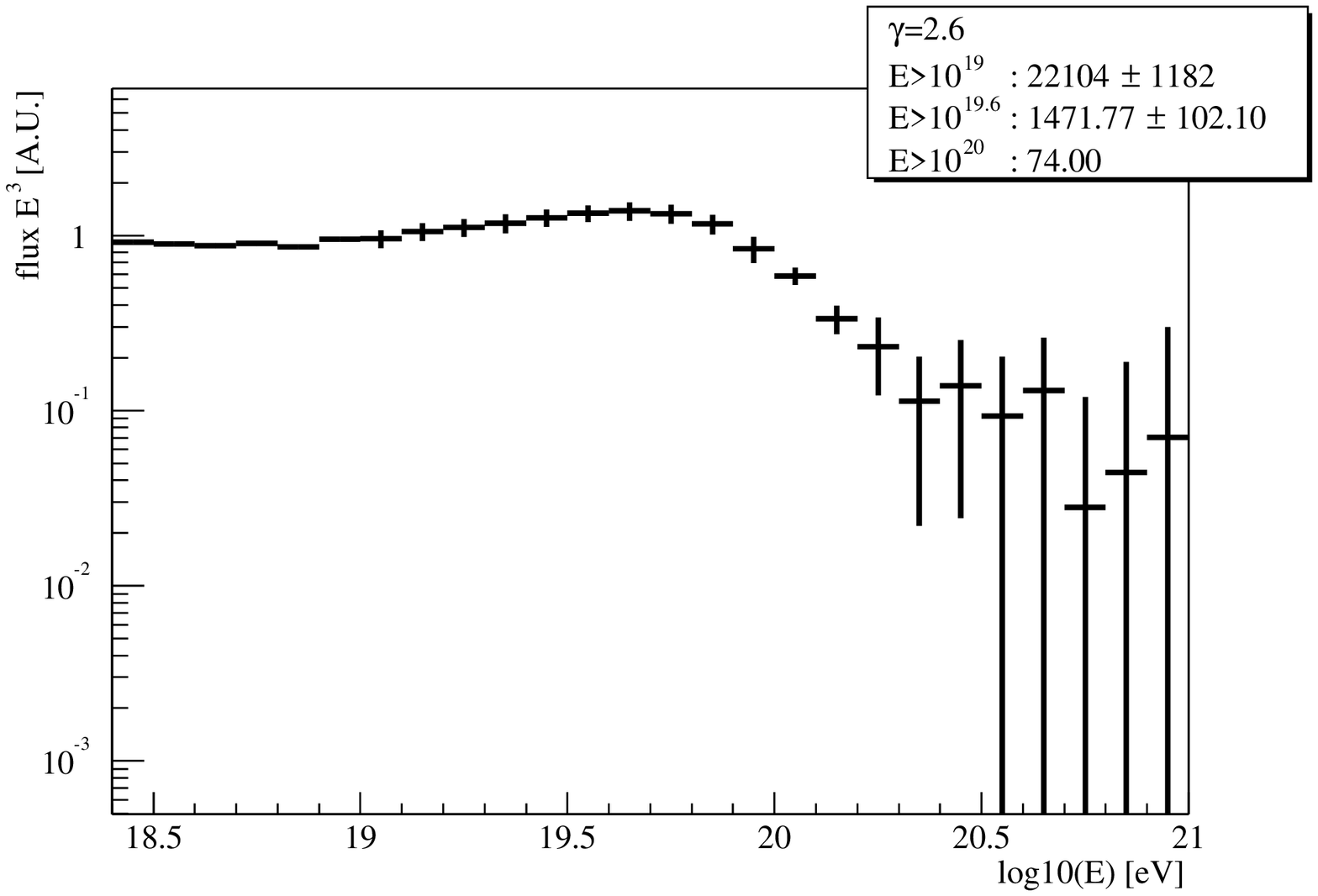}
\epsfxsize=6cm 
\epsfbox{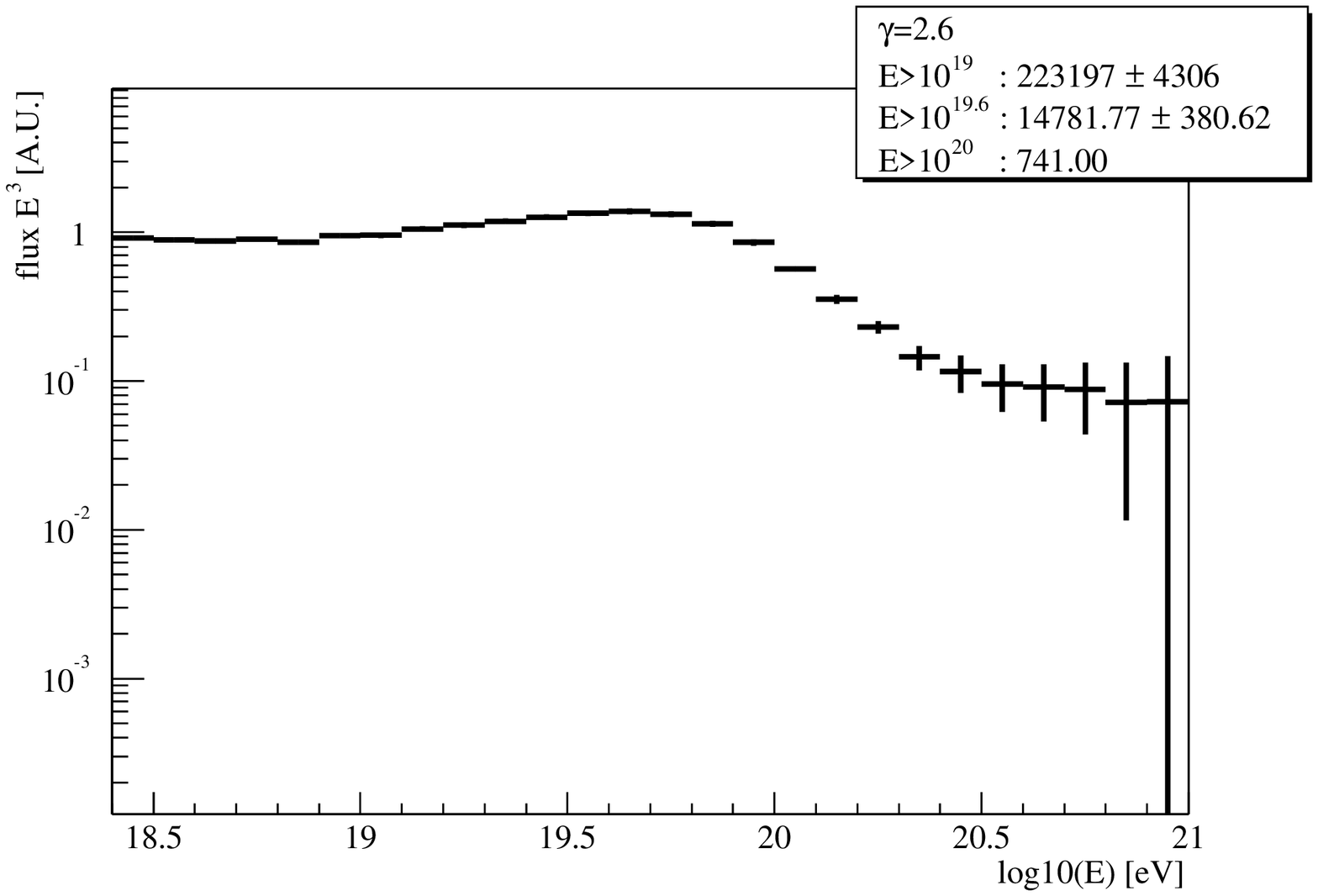}
\caption{Simulated spectra for the Auger (left) and EUSO (right) expected 
statistics of events as obtained in $^2$.}
\label{fig:future}
\end{figure}

From all this follows that the sharpness of the GZK feature depends on many
still unknown parameters. It is likely that the GZK feature is indeed rather
smooth, which implies a substantial number of events above $10^{20}$ eV. This 
is important for the assessment of the role of next generation UHECR 
experiments.
As an example, in Fig. \ref{fig:future} I plot the expected spectra of 
UHECRs for the Auger (left panel) and EUSO (right panel) statistics of events, 
as simulated in \cite{daniel1}.

The injection spectrum used in Fig. \ref{fig:future} is $E^{-2.6}$, which 
fits the low energy (AGASA and HiRes) data in the case that there is no 
evolution of the source luminosity with the redshift. 
Even a comparison by eye of the AGASA data with these 
simulated spectra shows that the Auger and EUSO statistics are needed in order 
to finally settle the issue of the presence of the GZK feature and its
detailed shape. 

\subsection{What are small scale anisotropies and large scale isotropy 
telling us?}

Despite the absence of any correlation between the arrival directions of UHECRs
and any local matter overdensity (e.g. the Galaxy or the local supercluster), 
some anisotropies on the scale of a few degrees have been found in the AGASA 
data \cite{agasaclusters} in the form of doublets and triplets of events. For 
a truly diffuse distribution of sources these multiplets should appear only as
a result of chance coincidences. This possibility appears to be disfavored, 
although some discussion is still ongoing \cite{finley2003} mainly concerning 
the way the data are treated for the analysis of the statistical significance.
In Hires data there seems to be no indication of such anisotropies, which is
not surprising for two reasons: 1) for the data obtained in the {\it mono} 
regime, the directions of arrival are obtained in the form of ellipses for 
which the definition of clusters of events is problematic. The data collected
in the {\it stereo} mode on the other hand are fewer; 2) the acceptance
of HiRes is energy dependent in the energy region above $10^{19}$ eV. This
means that the number of events needs to be corrected in order to obtain the
spectrum of UHECRs, but this correction cannot account for the directions of
arrival of the events that the correction is made for. 

Clustering of events on small angular scales is the signature that UHECRs are
accelerated at astrophysical sources. Moreover, the number of multiplets with 
different multiplicities is an index of the density of sources, which makes 
small scale anisotropies a very important tool to have the first important 
clue to the nature of the sources. 

Some attempts to estimate the number of sources of UHECRs in our cosmic
neighborhood from the small scale anisotropies found by AGASA have been carried
out, adopting both semi-analytical and numerical approaches. An analytical
tool to evaluate the chance coincidence probability for arbitrary statistics
of events was proposed in \cite{tom}. A rigorous analysis of the clusters
of events and of their energy dependence was given in \cite{will}. 
In \cite{dubovsky} the authors use an analytical method to estimate the 
density of the sources of UHECRs restricting their attention to the 14 events 
with energy above $10^{20}$ eV with one doublet. They obtain a rather 
uncertain estimate centered around $6\times 10^{-3}~\rm Mpc^{-3}$. In 
\cite{fodor} the energy losses are introduced through a function, derived 
numerically, that provides the probability of arrival of a particle from a 
source at a given distance.
Again, only events above $10^{20}$ eV are considered, therefore the analysis
is based upon one doublet of events out of 14 events. This causes extremely
large uncertainties in the estimate of the source density, found to be
$180^{+2730}_{-165} \times 10^{-3}~\rm Mpc^{-3}$. No account of the 
statistical errors in the energy determination nor of the declination 
dependence of the acceptance of the experimental apparata is included in 
all these investigations. A complete calculation of the number density and 
luminosity of the sources of UHECRs with a full numerical simulation of the 
propagation was carried out in \cite{daniel2}. Such a calculation also 
accounted for the statistical errors in the energy determination and the
declination dependence of the acceptance of the experiments involved.

\begin{figure}[ht]
\epsfxsize=12cm   %width of figure - will enlarge/reduce the figures
\epsfbox{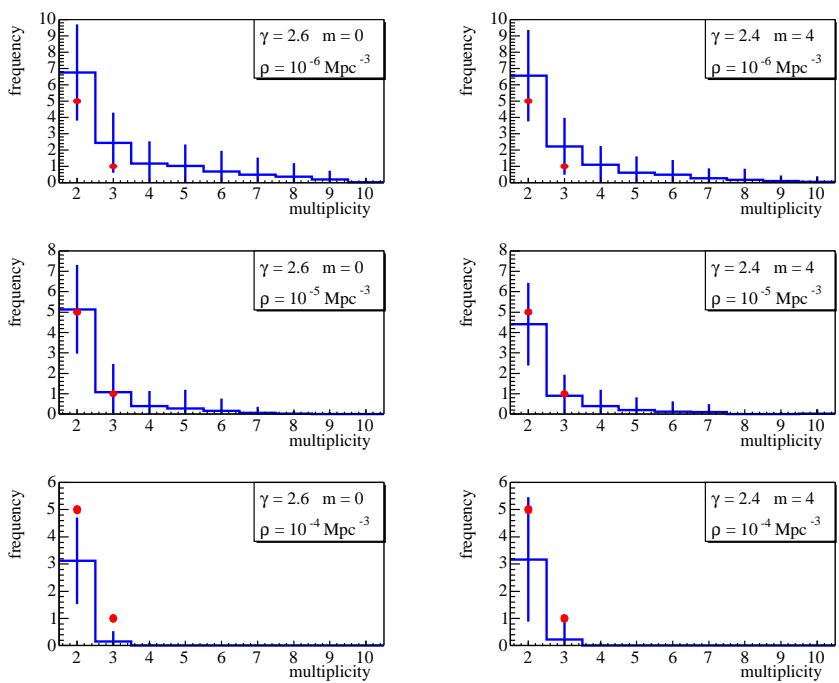}
\caption{Small scale anisotropies simulated in $^{13}$ for the AGASA 
statistics. The thick dots represent the number of doublets and triplets
actually observed.}
\label{fig:small}
\end{figure}

In Fig. \ref{fig:small} I plot the average frequency of occurrence of 
multiplets with different multiplicities using the AGASA statistics of 
events, as obtained in \cite{daniel2}. The thick dots represent the number 
of doublets and triplets observed by AGASA. 
The panels refer to source density $10^{-6}~\rm Mpc^{-3}$,
$10^{-5}~\rm Mpc^{-3}$, $10^{-4}~\rm Mpc^{-3}$ from top to bottom, and
are obtained for an injection spectrum $E^{-2.6}$ with no redshift evolution 
(left panels) and injection spectrum $E^{-2.4}$ with redshift evolution 
$\propto (1+z)^4$ (right panels). The error bars are obtained by simulating 
many realizations of the source distribution at fixed average source density.
The best fit seems to be obtained for a source density $n_s \approx 
10^{-5}~\rm Mpc^{-3}$, although the error bars are large due to the limited 
statistics of clustered events in AGASA. In \cite{daniel2} the simulation 
was extended to mock the statistics expected for Auger and EUSO. 

\begin{figure}[ht]
\epsfxsize=6cm   %width of figure - will enlarge/reduce the figures
\epsfbox{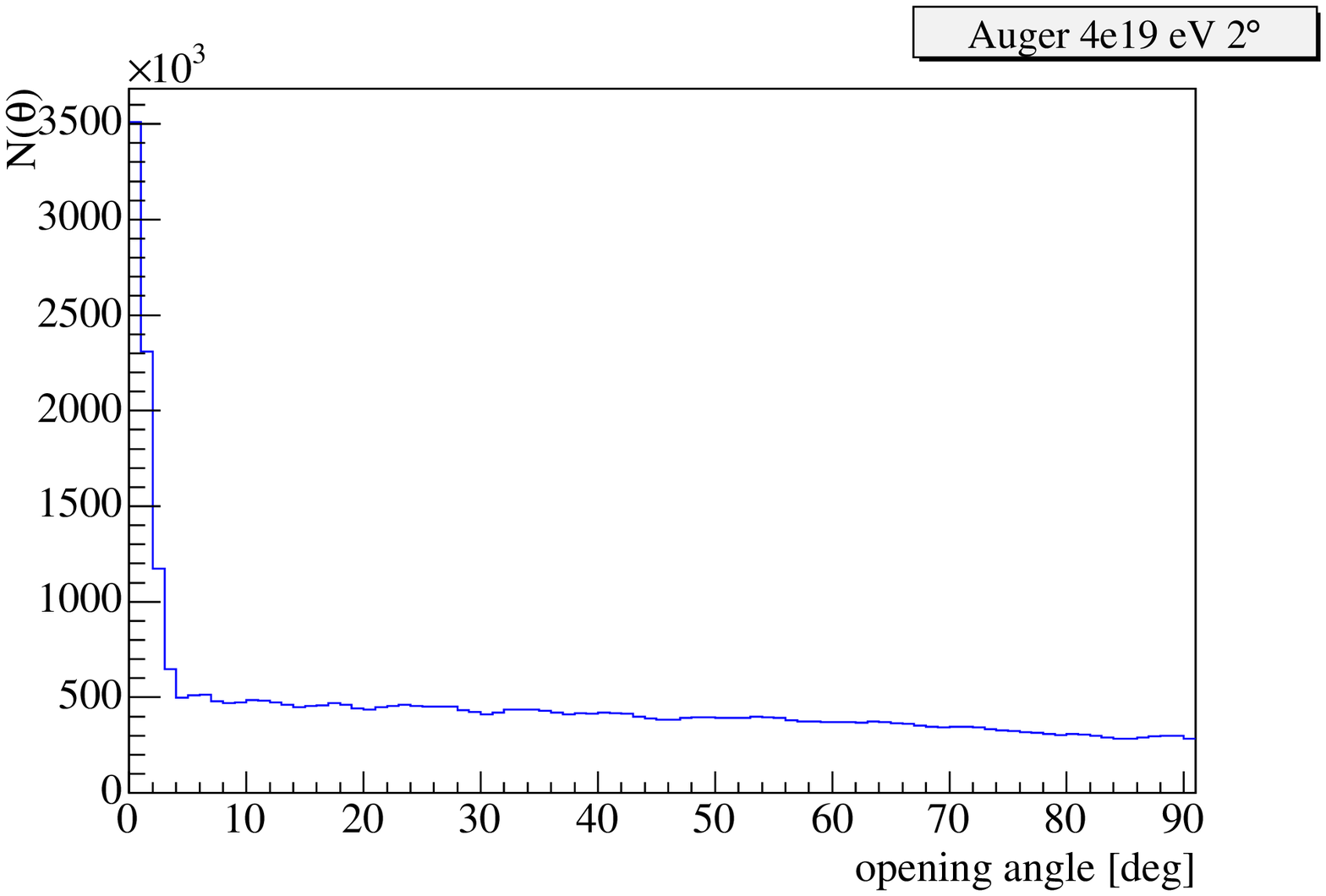}
\epsfxsize=6cm   %width of figure - will enlarge/reduce the figures
\epsfbox{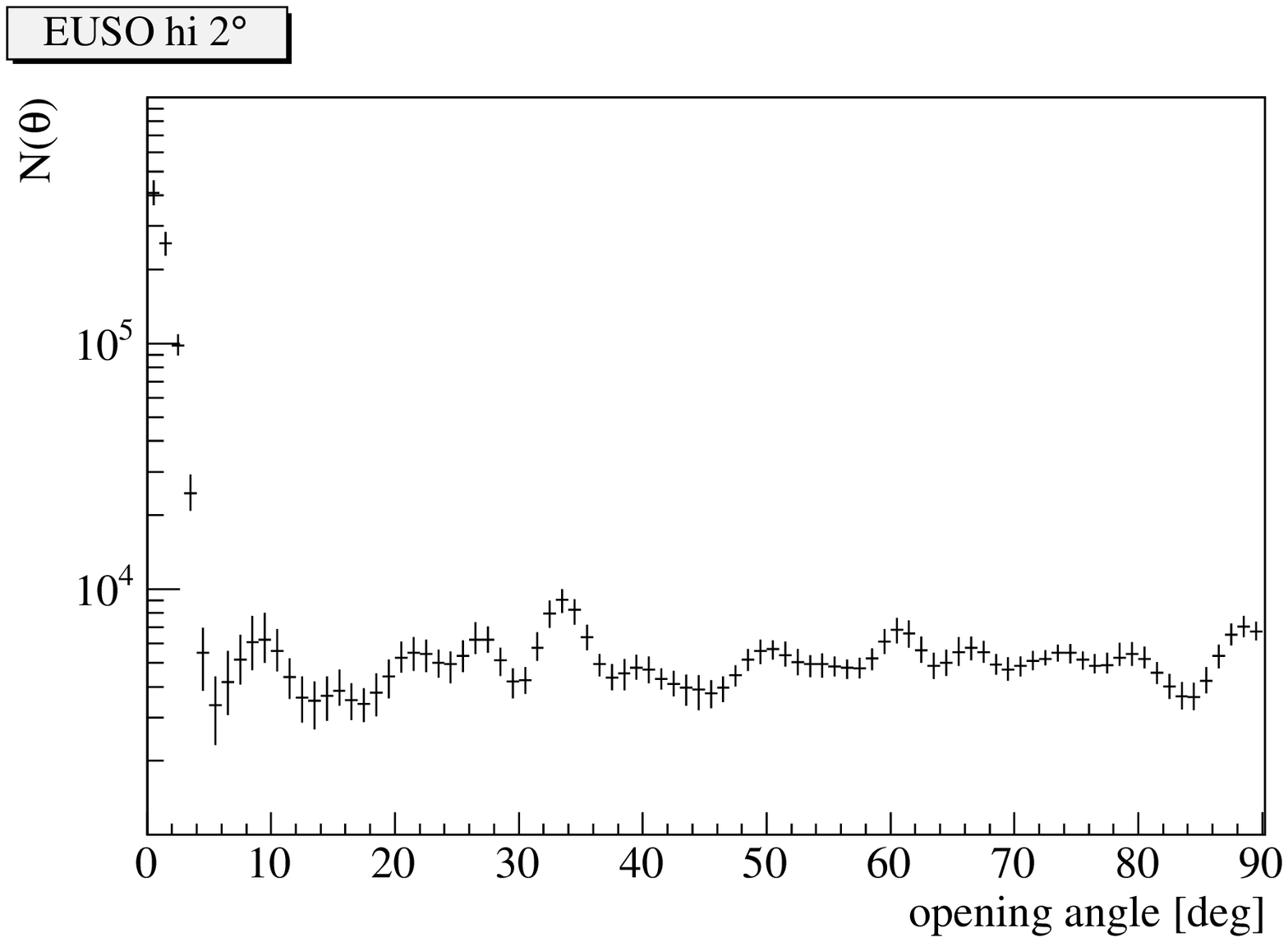}
\caption{Two point correlation function for the simulated Auger (left)
and EUSO (right) data, as obtained in $^{13}$ for an angular resolution 
of $2$ degrees. The Auger data refer to energies above $4\times 10^{19}$
eV, while the EUSO data refer to energies larger than $10^{20}$ eV, where
the acceptance is expected to be independent of energy.}
\label{fig:small2pt}
\end{figure}

In Fig. \ref{fig:small2pt} I plot the two-point correlation funtion for 
the simulated data of Auger (left panel) and EUSO (right panel) \cite{daniel2}.
The very pronounced peak at small angular separation is the evidence for 
clusters of events, while an approximately flat line would be expected in 
the case of perfect isotropy down to small scales. It is therefore clear that 
Auger and EUSO will definitely pin down the density of the sources of UHECRs 
(The figures were obtained for a source density of $10^{-5}~\rm Mpc^{-3}$). 
The wiggles in the EUSO two-point correlation function appear because of 
statistical fluctuations but their average amplitude is related to the number
density of sources \cite{daniel2}. These features start to appear when the 
exposure of the experiment gets close to the so-called {\it critical 
exposure}, $\sigma_c = 42000~ \rm (n_s/10^{-5}~\rm Mpc^{-3})~ km^2~sr~yr$,
defined as the exposure that allows at least one event with energy above 
$10^{20}$ eV to reach the detector from each single source within a
distance equal to the loss length of these particles. 

On large scales there is no evidence for anisotropies in the data. It 
has been argued that since there is a clear inhomogeneity in the nearby 
distribution of galaxies (local supercluster), the complete isotropy
at energies above $4\times 10^{19}$ eV would suggest the presence of a
strong local magnetic field \cite{sigl1}. In fact, as discussed at 
length in \cite{daniel1,daniel2}, particles with energy around 
$4\times 10^{19}$ can reach us from distances of $800-1000$ Mpc, therefore 
the local inhomogeneous distribution of sources does not reflect into the 
anisotropy of the arrival directions at this energy (to some extent, 
this appears to be the conclusion reached later in \cite{sigl2}).
In other words, at given energy, the observed anisotropy would reflect the 
inhomogeneity of the universe on scales comparable with the loss length at 
that energy.

\subsection{What is the role of intergalactic magnetic fields on the 
propagation of UHECRs?}

Our knowledge of the intergalactic magnetic field is very poor and in 
order to give a fair answer to this question all possibilities should 
be investigated, provided no observational bound is violated.
Magnetic fields of $0.1-1~\mu G$ have been found in virialized 
structures, such as galaxies and clusters of galaxies. These sources 
however occupy a fraction $\sim 10^{-5}$ of the volume of the universe
and do not play an important role for the propagation of the bulk of 
UHECRs. The magnetic fields that we are most interested in are therefore
the fields in the voids and in non-virialized large scale structures, 
that occupy most of the volume of the universe. If the magnetic field in 
the universe is the result of the pollution of astrophysical sources,
these regions are expected to be not much magnetized: most field should
be where sources are, for the same reason why gas is mostly where sources are.
If however the field has a cosmological origin, then even the voids 
and filamentary regions may have appreciable magnetization. Observationally,
the most stringent limits come from Faraday rotation measurements (FRM): 
these limits depend on assumptions on the topology of the field and on the
density of matter along a line of sight. Current limits vary wildly depending
on the assumptions adopted, the most severe limit being $B<10^{-9}-10^{-10}$
G (e.g. Kronberg \cite{kronberg}, but correcting his limit for the fact that
$\Omega_b$, the baryon fraction in the universe, is not unity. See 
\cite{burles} for a discussion). Charged particles in a turbulent field
perform a random walk. The departure of the trajectory from a straight line 
stays smaller than some given angular size $\theta_{exp}$ if 
$$B < 10^{-10} \left(\frac{E}{4\times 10^{19}eV}\right)
\left(\frac{\theta_{exp}}{2^o}\right)\left(\frac{D}{1000\rm Mpc}\right)^{-1/2} 
\left(\frac{L_c}{1\rm Mpc}\right)^{-1/2} ~ \rm Gauss.$$
Here we can interpret the angle $\theta_{exp}$ as the angular resolution of
an experiment for the detection of UHECRs.
This expression suggests that particles with energy $\sim 4\times 10^{19}$ eV
move in approximate straight line propagation provided the field on the scale 
specified by the distance $D$ remains smaller than $\sim 10^{-10}$ G. 
This result is confirmed by recent numerical simulations of the propagation 
of UHECRs in the magnetized large scale structures of the universe 
\cite{grasso}. For lower energy particles the deflection may become large and 
the propagation may approach the diffusive regime (see \cite{stanev} for a 
discussion). 

The limit imposed by FRM becomes even less stringent on the field in local
structures such as the local supercluster. In this case however one can use
some physical insights in order to get a hint on the strength of the field:
at the present cosmic time, clusters of galaxies are the largest virialized 
regions in the universe. Larger structures, such as superclusters, did not 
get to that stage as yet, and are therefore expected to have smaller velocity 
dispersion and smaller magnetic fields. The magnetic fields measured in 
clusters of galaxies (in fact in only some of them) is of order $0.1-\rm few
\mu G$, roughly $10-100$ times smaller than the equipartition fields 
\footnote{One should keep in mind that the concept of equipartition field 
here is different from the one typically used in radio astronomy, in which 
the equipartition is meant with the radiating relativistic particles. The
latter field is clearly quite smaller than the former.} (despite 
the fact that these are virialized structures). In general, if $R_{size}$ is 
the size of a structure and $M$ the mass in the volume $(4/3)\pi R_{size}^3$, 
the condition $\frac{B^2}{8\pi} \ll \rho v^2$ translates into
$$B \ll 8\pi \left(\frac{G}{3}\right)^{1/2} R_{size} (\Omega_b \Omega_m)^{1/2}
\rho_{cr} \approx 10^{-7} \left(\frac{R_{size}}{10~Mpc}\right) 
\left(\frac{h}{0.7} \right)^{3/2} G,$$
where $h$ is the dimensionless Hubble constant and I used $\Omega_b=
0.02 h^{-2}$ for the baryon fraction and $\Omega_m=0.3$ for the mass 
fraction. 

\subsection{Where and how are UHECRs accelerated?}

A first piece of the answer to this question may come from the measurement
of the spectral shape in the energy region where we expect to find the GZK 
feature and from the measurement of the chemical composition. For instance 
a weak or absent GZK feature, together with an appreciable fraction of gamma 
rays in the composition would point toward a top-down origin of UHECRs. In this
case it is difficult to envision how prominent small scale anisotropies may
arise. In the following, due to the limited space available I will not 
discuss these models any further (see \cite{batta} for a review of this class
of models). 

As discussed above, both the shape of the GZK feature and the presence of
small scale anisotropies should tell us about the source density and
luminosity. For sources which are continuous in time, as stressed above, the 
source density inferred from SSA is $\sim 10^{-5} Mpc^{-3}$, corresponding to 
a source luminosity $\sim 10^{42}~\rm erg~s^{-1}$ at energies above $10^{19}$
eV. 

A discussion of {\it 'Where'} may UHECRs be accelerated can be found in many 
excellent reviews \cite{angela,floyd,proth,blasirev} and the limited space 
available here does not allow me to add much to what is discussed there. 
In the following I will therefore concentrate on two issues concerning
the {\it 'How'} UHECRs are accelerated. More specifically I will spend a
few words on two subjects, namely non-linear shock acceleration at newtonian
shocks, and particle acceleration at relativistic shocks that are particularly
important for the acceleration of UHECRs. 

Most scenarios that have been proposed so far for the acceleration of UHECRs 
in astrophysical sources, with few exceptions, are related to particle 
acceleration at either newtonian or relativistic shock waves. It is therefore
important that we understand the details of this process in order to confront
observations. A typical assumption adopted to estimate the plausibility of a
class of sources is that the spectrum of accelerated particles is a power law
and the energy it contains is a small fraction of the total energy budget. 
It has been found in recent times that neither one of these assumptions need to
be valid. Numerical calculations suggest \cite{simula} that although the 
fraction of particles accelerated to suprathermal energies is in fact small, 
the energy channelled into these few particles may be a substantial part of 
the kinetic energy flux through the shock surface. 
This effect is due to the backreaction of the accelerated particles on the 
shock itself, that enhances the efficiency of acceleration instead of reducing 
it. When this happens, the spectrum of the accelerated particles is no longer 
a power law, as shown also in analytical calculations of the nonlinear 
backreaction \cite{malkov,blasi}.
More specifically, for strongly modified shocks the spectra are such that most 
energy is concentrated at $p_{max}$, the maximum momentum achievable at the 
shock, something which is at odds with the predictions of the linear theory, 
where the energetics of accelerated particles is typically dominated by the 
mildly relativistic particles. 
It is easy to envision the consequences that these findings may have for the 
acceleration of UHECRs, since in this case there is the constant need to 
channel as much energy as possible in the highest energy part of the spectrum
of a source.

In some candidate sources of UHECRs, such as gamma ray bursts, the acceleration
is postulated to occur at relativistic shock waves. The spectrum of accelerated
particles in these cases depends quite sensibly on the scattering
properties of the media upstream and downstream of the shock. In most cases
however, the spectra of accelerated particles are calculated in the assumption 
of small pitch angle scattering and are found to be power laws with a slope
that for ultra-relativistic shocks tends to be almost universal \cite{uni}, 
$\gamma\sim 2.2-2.4$. 
Recently in \cite{vietri_s} a new approach has been proposed that 
describes shock acceleration at shocks with arbitrary velocity (from newtonian
to relativistic) and arbitrary scattering properties of the medium. In 
\cite{blasi_s}, the approach was further developed and checked versus several 
test cases in both the newtonian and relativistic regimes. The approach allows 
one to determine analytically the spectrum of particles and the anisotropy in 
pitch angle even for those situations in which the assumption of small pitch 
angle scattering does not apply. 

Some general comments on shock acceleration may be made independently of the 
details: the acceleration of particles at ultra-relativistic shocks is likely
to proceed in two steps. During the first step, a factor $\sim \Gamma^2$ can be
gained by the particles ($\Gamma$ here is the Lorentz factor of the shock). 
After this first step, the distribution of particles upstream is beamed within 
a cone of aperture $1/\Gamma$ with respect to the direction of motion of the
shock and the energy gain is only of order unity in the successive shock
crossings. The acceleration at relativistic shocks requires some appreciable
level of turbulence in the downstream fluid: in such frame, the shock moves 
with a speed $\sim 1/3$, therefore the particles would not be able to reach 
the shock if the field is coherent on large scales. A recent review of shock 
acceleration at relativistic shocks can be found in \cite{ostro}.

\section{Conclusions}

I discussed some general issues related to the origin and propagation of 
UHECRs, trying to point out which future developments may contribute to 
improve our understanding of the current problems.

\end{document}